\documentclass{article}
\usepackage{graphicx}
\usepackage{amsmath}
\usepackage{amsfonts}
\usepackage{amssymb}
\newtheorem{theorem}{Theorem}

\newtheorem{definition}[theorem]{Definition}

\begin{document}

\title{Correlated Equilibria of Classical Strategic Games with Quantum Signals}
\author{Pierfrancesco La Mura\\Leipzig Graduate School of Management\\plamura@hhl.de\\comments welcome}
\maketitle
\begin{abstract}
Correlated equilibria are sometimes more efficient than the Nash equilibria of
a game without signals. We investigate whether the availability of quantum
signals in the context of a classical strategic game may allow the players to
achieve even better efficiency than in any correlated equilibrium with
classical signals, and find the answer to be positive.
\end{abstract}

\section{Introduction}

Until recently, the most notable connection between game theory and quantum
mechanics was the fact that von Neumann pioneered the mathematical foundations
of both fields [10,11]. In recent years a few proposals have appeared in the
literature, mostly coming from the theoretical physics community, aimed at
finding a common ground for the two theories. Several recent papers have
investigated the connections between games with classical mixed strategies and
games with quantum strategies [3,8]. Generally, these results have had limited
resonance in the game theory community. Unlike the vast majority of real-life
strategic situations, in which the available strategies are predetermined and
classical, a game with quantum strategies needs to be especially designed and
implemented in order to be played. Moreover, it was soon pointed out that any
such quantum game can be simulated by a suitably designed classical game (in
which the players, say, communicate via telephone their choice of quantum
strategy to a center), and hence interest in the field of quantum games has
been so far relatively modest.

The aim of the present paper is also to bring together ideas from game theory
and quantum mechanics, but following a significantly different approach, which
is closest in spirit to [6] and [7]. We take the view that in many games the
strategies available to the players are predetermined and classical, but
following Aumann [1,2] we postulate the availability of a correlating device
able to send payoff-irrelevant private signals to the players before they play
the game. Unlike Aumann's device, which can only send classical information,
we also allow for devices able to send sequences of individual quantum signals
(qubits) to the players. Today similar devices are amply available, as many
practical implementations have been designed and demonstrated. A qubit may be
variously encoded in the properties of a physical carrier, including the spin
of a neutron, or the polarization of a photon.

Correlated equilibria are often more efficient than the Nash equilibria of a
game without signals. This is because the correlation of the signals received
by the players may allow them to coordinate towards Pareto-superior expected
payoff combinations. We investigate whether allowing the use of quantum
signals may allow the players to achieve even better coordination than is
possible in any classical correlated equilibrium, and find the answer to be
positive. This has some implications on the significance of several well-known
impossibility results [9,5,12], which will need to be re-evaluated if quantum
correlation is allowed.

The main contribution of this paper is to introduce a novel notion of
correlated equilibrium in which the signals sent to the players are
represented by quantum bits, and compare its properties with those of
classical correlated equilibrium. Classical correlated equilibrium deals with
information states which take the form of events in a suitably defined Boolean
algebra. By contrast, the information states induced by quantum signals
generally do not correspond to a Boolean algebra, but conform to a more
general logic calculus in which the distributive laws of propositional logics
appear in a weakened form.

Because of its unique features, the field of quantum information theory is
currently an object of intense study, together with the rising fields of
quantum computation, encryption and communication. The most distinguishing
feature of quantum theory, which sets it clearly apart from classical
mechanics, is the phenomenon of entanglement [4], by which the observed states
of two particles may exibit correlations which cannot be rationalized in terms
of classical probabilistic behavior. If two entangled particles are kept
isolated from their local environment they will remain in a correlated state
until some operation is performed on one of them (say, the spin is observed),
in which case the states of both particles will change in a predictably
correlated way, even if they have been meanwhile separated in space and time.
This phenomenon is non-local: the states of the two particles are jointly
affected, even though only one of them is observed.

\section{An example}

Consider the following game with imperfect information. Two players are
randomly drawn from an i.i.d., uniform distribution over a set of three types,
$\{A,$ $B,C\}.$ They privately observe their own type, then simultaneously
decide whether to say ``yes'' or ``no''. If two players of the same type
agree, i.e. they both say yes or they both say no, each of them receives a
payoff of $-900.$ If two players of different types agree they both receive a
payoff of $9.$ If the two players disagree they both get zero. Due to the
catastrophic consequences of agreeing when of the same type, there are no
efficient Nash equilibria which give positive probability to those events. All
the possible pure strategy combinations which do not lead to catastrophes are
(denoting ``no'' by $0$ and ``yes'' by $1$)

\bigskip%

\begin{tabular}
[c]{llllllll}%
player 1 & A & B & C & player2 & A & B & C\\
a & 0 & 0 & 0 &  & 1 & 1 & 1\\
b & 0 & 0 & 1 &  & 1 & 1 & 0\\
c & 0 & 1 & 0 &  & 1 & 0 & 1\\
d & 0 & 1 & 1 &  & 1 & 0 & 0\\
e & 1 & 0 & 0 &  & 0 & 1 & 1\\
f & 1 & 0 & 1 &  & 0 & 1 & 0\\
g & 1 & 1 & 0 &  & 0 & 0 & 1\\
h & 1 & 1 & 1 &  & 0 & 0 & 0
\end{tabular}

\bigskip

Strategy combinations $a$ and $h$ lead to an expected payoff of zero. Any of
the other strategy combinations leads to an expected payoff of $4$. This game
has $175$ Nash equilibria, $8$ of which are in pure strategies and correspond
to strategy combinations $a,b,...,h$ above. The only efficient equilibria are
in pure strategies, and correspond to strategy combinations $b,c,...,g$.
Furthermore, $(4,4)$ are not only the efficient Nash equilibrium payoffs, but
also the efficient correlated equilibrium payoffs. In fact, it is well known
that the set of correlated equilibrium payoffs which can be obtained with
public signals is the convex hull of the set of Nash equilibrium payoffs. But
this is a game of pure coordination, and hence any correlated equilibrium with
private signals is still an equilibrium if the signals are made public (as the
players can always ignore the extra information). It follows that $(4,4)$ are
also the efficient correlated equilibrium payoffs.

\bigskip

Now suppose that each player receives one half of a fully entangled qubit
pair, and that three different measurements, $x,y$ and $z,$ are possible (say,
spin measurements along three commonly known orthogonal axes). Then the
following equilibrium improves on $(4,4).$ If of type $A,$ choose measurement
$x.$ Say ``yes'' if and only if the result is ``up'', and say ``no''
otherwise. Behave analogously, if of type $B,$ after applying measurement $y,$
and if of type $C$ after applying measurement $z.$ It follows from the rules
of quantum mechanics that the conditional probabilities of agreement,
depending on the combination of types, are those represented in the following table.

\bigskip%

\begin{tabular}
[c]{llll}%
& A & B & C\\
A & 0 & 3/4 & 3/4\\
B & 3/4 & 0 & 3/4\\
C & 3/4 & 3/4 & 0
\end{tabular}

\bigskip

It is easily checked that the strategy profile described above is an
equilibrium, yielding an expected payoff of $6/9\ast3/4\ast9=9/2>4$ to both players.

\bigskip

Some remarks are in order at this point.

\begin{itemize}
\item  The game is constructed in order to take advantage of the violation of
one of Bell's inequalities by the entangled qubit pair. No distribution of
classical signals can induce the conditional probabilities of agreement in the
table above, and this the reason why no classical correlated equilibrium can
achieve the same expected payoffs.

\item  In general, better coordination than in classical correlated
equilibrium may ensue due to the rich patterns of positive and negative
correlation which are allowed by the rules of quantum mechanics.

\item  Note that no actual information is exchanged between the two players.
From the point of view of each player the probability that the opponent is of
any given type is still $1/3,$ and the probability that he will say ``yes'' is
still $1/2,$ even after performing the observation. Yet, the strategies of the
two players have become effectively entangled, by making them contingent on
the choice of measurement and on the observed state of the qubit pair.

\item  The correlated equilibrium described above makes use of a single
entangled qubit pair, but possibly even better coordination may be achieved by
transmitting finite sequences of entangled qubit pairs (q-bytes).

\item  Finally, note that in contrast with the classical case here observing
is explicitly modelled as an act. This is necessary in the quantum-theoretic
framework, where any observation intrinsically affects the state of the
observed system.
\end{itemize}

\section{Game-theoretic setup and a formal definition}

The example described in the previous section can be conveniently modelled as
a Bayesian game. In this section we introduce some standard game-theoretic
notation and definitions, and then give a formal definition of quantum
correlated equilibrium. We start with the definition of a strategic game.

\begin{definition}
A \emph{strategic game} consists of a finite set $N$ (the set of
\emph{players}), and for each player $i\in N$

\begin{itemize}
\item  a set $A_{i}$ (the set of \emph{actions} available to $i$)

\item  a function $u_{i}:A\rightarrow\mathbb{R},$ where $A=\times_{j\in
N}A_{j}$ (the \emph{payoff} of player $i$).
\end{itemize}
\end{definition}

Denote by $\Sigma_{i}$ the set of probability measures over $A_{i}.$ These are
player $i$'s \emph{mixed strategies}. We use the suffix $-i$ to denote all
players except $i.$ For instance, a profile $a_{-i}\in A_{-i}$ specifies an
action for all players except $i.$ A Nash equilibrium is a combination of
probabilistic decisions (mixed strategies), one for each player, such that no
single player can get a better expected payoff by choosing a different
strategy. 

\begin{definition}
A (mixed-strategy) Nash equilibrium of a strategic game $(N,(A_{i}),(u_{i}))$
is a vector $(\pi_{1},...,\pi_{n})$, with $\pi_{i}\in\Sigma_{i}$ for all $i\in
N,$ such that
\[
\sum_{a\in A}\pi_{i}(a_{i})\pi_{-i}(a_{-i})u_{i}(a_{-i},a_{i})\geq\sum_{a\in
A}\rho_{i}(a_{i})\pi_{-i}(a_{-i})u_{i}(a_{-i},a_{i})
\]

for all $\rho_{i}\in\Sigma_{i}$ and for all $i\in N.$
\end{definition}

Pure-strategy Nash equilibria are those which only involve degenerate mixed
strategies. In a Bayesian game, each player can make its choice of available
action contingent on the observation of a private signal.

\begin{definition}
A \emph{Bayesian game} (in strategic form) consists of:

\begin{itemize}
\item  a finite set $N$ (the set of players)

\item  a finite set $\Omega$ (the set of \emph{states})
\end{itemize}

and for each player $i\in N$

\begin{itemize}
\item  a set $A_{i}$ (the set of actions available to $i$)

\item  a finite set $T_{i}$ (the set of \emph{signals} that may be observed by
player $i$) and a function $\tau_{i}:\Omega\rightarrow T_{i}$ (the
\emph{signal function} of player $i$)

\item  a probability measure $p_{i}$ on $\Omega$ (the \emph{prior belief} of
player $i$) for which $p_{i}(\tau_{i}^{-1}(t_{i}))>0$ for all $t_{i}\in T_{i}$

\item  a function $u_{i}:A\times\Omega\rightarrow\mathbb{R},$ where
$A=\times_{j\in N}A_{j}$ (the payoff of player $i$).
\end{itemize}
\end{definition}

Any Bayesian game can be reduced to a suitably defined strategic game. The
Nash equilibria of a Bayesian game are defined as the Nash equilibria of the
corresponding strategic game.

\begin{definition}
A \emph{Nash equilibrium} of a Bayesian game $(N,\Omega,(A_{i}),(T_{i}%
),(\tau_{i}),(p_{i}),(u_{i}))$ is a Nash equilibrium of the strategic game
defined as follows.

\begin{itemize}
\item  The set of players is the set of pairs $(i,t_{i})$ for $i\in N$ and
$t_{i}\in T_{i}.$

\item  The set of actions of each player $(i,t_{i})$ is $A_{i}.$

\item  The payoff function $u_{(i,t_{i})}$ of each player $(i,t_{i})$ is given
by $u_{(i,t_{i})}(a)=E[u_{i}(a,\omega)|t_{i}]$ (where $E[$ $\;]$ denotes the
expectation operator)
\end{itemize}
\end{definition}

While a Nash equilibrium captures the notion of simultaneous, independent
decisions, in a correlated equilibrium the players can make their choice of
strategy conditional on some privately observed classical signals. Even though
the signals are payoff-irrelevant, they influence the expected payoff by
changing the players' information states.

\begin{definition}
A \emph{correlated equilibrium} of a strategic game $(N,(A_{i}),(u_{i}))$
consists of

\begin{itemize}
\item  a finite probability space $(\Omega,\pi)$ ($\Omega$ is the set of
states and $\pi$ is a probability measure on $\Omega$)

\item  for each player $i\in N$ a partition $\mathcal{P}_{i}$ of $\Omega$
(player $i$'s \emph{information partition})

\item  for each player $i\in N$ a function $\sigma_{i}:\Omega\rightarrow
A_{i}$ with $\sigma_{i}(\omega)=\sigma_{i}(\omega^{\prime})$ whenever
$\omega\in P_{i}$ and $\omega^{\prime}\in P_{i}$ for some $P_{i}\in
\mathcal{P}_{i}$ such that%
\[
\sum_{\omega\in\Omega}\pi(\omega)u_{i}(\sigma_{-i}(\omega),\sigma_{i}%
(\omega))\geq\sum_{\omega\in\Omega}\pi(\omega)u_{i}(\sigma_{-i}(\omega
),\rho_{i}(\omega))
\]

for any other function $\rho_{i}:\Omega\rightarrow M_{i}\times A_{i}$ such
that $\rho_{i}(\omega)=\rho_{i}(\omega^{\prime})$ whenever $\omega\in P_{i}$
and $\omega^{\prime}\in P_{i}$ for some $P_{i}\in\mathcal{P}_{i}.$
\end{itemize}
\end{definition}

As for the case of Nash equilibria, the correlated equilibria of a Bayesian
game are simply the correlated equilibria of the associated strategic game.

\bigskip

We are now in the position to give a formal definition of quantum correlated
equilibrium in a strategic game. The definition is similar to that of
correlated equilibrium, but now the choice of measurement is explicitly
modelled as an act.

\begin{definition}
A \emph{quantum correlated equilibrium }of a strategic game $(N,(A_{i}%
),(u_{i}))$ consists of

\begin{itemize}
\item  a finite probability space $(\Omega,\pi$)

\item  for each player $i\in N,$ a set $M_{i}$ (a subset of the set of
Hermitian operators on a complex, separable Hilbert space) representing the
available measurements

\item  for each player $i\in N,$ and for each $m_{i}\in M_{i},$ a partition
$\mathcal{P}_{i}^{m_{i}}$ of $\Omega$ representing player $i$'s information

\item  for each player $i\in N,$ a function $\sigma_{i}:\Omega\rightarrow
M_{i}\times A_{i},$ with $\sigma_{i}^{a}(\omega)=\sigma_{i}^{a}(\omega
^{\prime})$ whenever $\sigma_{i}^{m}(\omega)=\sigma_{i}^{m}(\omega^{\prime})$
and $\omega\in P_{i}$ and $\omega^{\prime}\in P_{i}$ for some $P_{i}%
\in\mathcal{P}_{i}^{\sigma_{i}^{m}(\omega)}$ (a \emph{strategy} of player
$i$), such that for any other strategy $\rho_{i}$%
\[
\sum_{\omega\in\Omega}\pi(\omega)u_{i}(\sigma_{-i}^{a}(\omega),\sigma_{i}%
^{a}(\omega))\geq\sum_{\omega\in\Omega}\pi(\omega)u_{i}(\sigma_{-i}^{a}%
(\omega),\rho_{i}^{a}(\omega)),
\]

where $\sigma_{i}^{a}(\omega)$ and $\sigma_{i}^{m}(\omega)$ denote,
respectively, the measurement and the action selected by strategy $\sigma_{i}$
at state $\omega.$
\end{itemize}
\end{definition}

Observe that the choice of measurement does not appear in the payoff functions
$u_{i}$, but influences the expected payoff via the players' information
partitions. The definition of quantum correlated equilibrium insures that
there is no profitable unilateral strategy deviation after a choice of
measurement has been performed and its result observed by a player, and also
insures that there is no other choice of measurement which may lead to higher
expected payoffs. Any correlated equilibrium is a quantum correlated
equilibrium (with just a single measurement available to each player). The
converse is not true in general, unless all the available measurements commute.

\bigskip

Note that a quantum correlated equilibrium always exists. In fact, if all the
players choose to ignore the results of measurement and play a Nash
equilibrium of the classical game this is always a quantum correlated
equilibrium according to the above definition. More remarkably, the set of
payoff combinations sustained by quantum correlated equilibria can be strictly
larger than the set of classical correlated equilibrium payoffs, as the
example in the previous section demonstrates.

\bigskip

It is not difficult to check that the equilibrium with two entangled qubits
described in the example satisfies the definition of quantum correlated
equilibrium. In the example, the set of states can be identified with the set
of all combinations of type, measurement and action by each player. The
information partition of player $i$ captures which measurement $(x,y$ or $z$)
was chosen by $i,$ and its outcome (``up'' or ``down''). A strategy specifies
which of the three measurements is chosen, and restricts the subsequent choice
of classical action (announcing ``yes'' or ``not'') to depend only on the type
of the player, on the choice of measurement and on its outcome. In
equilibrium, with probability $\left(  6/9\right)  \ast\left(  3/4\right)  $
the joint strategy yields a payoff of $9,$ and with the complementary
probability yields a payoff of zero. Note that the equilibrium is strict: any
other choice of action (for any given type, measurement and observation), as
well as any other choice of measurement, would lead to a stricly lower
expected payoff.

\section{Discussion}

We introduced a novel notion of correlated equilibrium which takes into
explicit account the fact that any information received by the players prior
to playing the game must be embodied in a physical carrier, and showed that
the availability of a correlating device able to send quantum signals to the
players may allow for equilibria which are more efficient than any classical
equilibrium. One of the two approaches in the literature which comes closest
to ours is the one in [7], which defines a class of games (`Quantum
Correlation Games', or QCG) in which quantum correlation is exploited to yield
more efficient equilibria if entangled quantum signals are available to the
players. Yet, in QCGs the only available actions are choices of measurement,
which makes QCGs a rather limited class of games. Unlike [7], we proposed an
equilibrium notion which applies to any classical strategic-form game,
provided that quantum signals can be sent to the players and observed prior to
their choice of classical action. Also, unlike [7], in our approach departures
from classical correlated equilibrium are directly related to violations of
Bell's inequalities.

The other approach which comes closest to ours is the one in [6], which also
postulates the availability of quantum signals and demonstrates how
entanglement can induce correlation in the players' strategies but fails to
relate these ideas with the notion of correlated equilibrium, and to realize
that quantum correlated equilibria can be strictly more efficient than
classical correlated equilibria. In particular, in the game used in [6] to
demonstrate quantum correlation, correlated equilibria based on classical
private signals would be equally efficient. Unlike [6], we esplicitly related
the notion of entanglement with that of correlated equilibrium in strategic
games, and showed that in some games quantum correlated equilibria can be
strictly more efficient than classical ones.

\section{References}

$\;\;$

[1] R.J. Aumann. Subjectivity and correlation in randomized strategies.
Journal of Mathematical Economics, 1, 1974.

[2] R.J. Aumann. Correlated equilibrium as an expression of Bayesian
rationality. Econometrica , 55, 1987.

[3] J. Eisert, M. Wilkens, and M. Lewenstein. ``Quantum games and quantum
strategies'', Physical Review Letters 83, 3077-3080, 1999.

[4] A. Einstein, B. Podolsky, and N. Rosen. ``Can Quantum-Mechanical
Description of Physical Reality be Considered Complete?'', Physical Review 47,
777-780, 1935.

[5] A. Gibbard. ``Manipulation of voting schemes: a general result'',
Econometrica, 41, 587-601, 1973.

[6] B. A. Huberman and T. Hogg, ``Quantum Solution of Coordination Problems'',
arXiv:quant-ph/0306112, July 2003.

[7] A. Iqbal and S. Weigert, ``Quantum Correlation Games'',
arXiv:quant-ph/0306176, Jun 2003.

[8] D. A. Meyer, ``Quantum strategies'', Physical Review Letters 82,
1052--1055, 1999.

[9] R. B. Myerson and M. A. Satterthwaite. ``Efficient mechanisms for
bilateral trading'', Journal of Economic Theory, 29:265--281, 1983.

[10] J. von Neumann and O. Morgenstern. Theory of Games and Economic Behavior.
Princeton University Press, 1944.

[11] J. von Neumann. Mathematical Foundations of Quantum Mechanics, Princeton
University Press, 1955.

[12] Mark A. Satterthwaite, ``Strategy-Proofness and Arrow's Conditions:
Existence and Correspondence Theorem for Voting Procedures and Social Welfare
Functions'', Journal of Economic Theory 10, 187-217, 1975.
\end{document}